\documentclass[aps]{revtex4}
\usepackage{graphicx}
\usepackage[figuresright]{rotating}
\usepackage{amsmath}
\begin{document}
\begin{flushright}
UCRL-JRNL-226480
\end{flushright}
\baselineskip=18pt

%\begin{titlepage}
\title{\bf No-core shell model for $ A = 47$ and $ A = 49$}
\author{A.~G.~Negoita$^{a,c}$, J.~P.~Vary $^{a,b}$, S.~Stoica$^{c}~$\\
\it $^a$
Department of Physics and Astronomy, Iowa State University, Ames, IA  50011\\
\it $^b$
Lawrence Livermore National Laboratory, Livermore, CA 94551\\
\it $^c$
Horia Hulubei National Institute for Physics and Nuclear\\
Engineering, P.O. Box MG-6, 76900 Bucharest-Magurele, Romania  }

\maketitle
\vskip5mm

%\vspace{0.5cm}
%\begin{abstract}
\noindent
{\bf Abstract}

We apply the no-core shell model to the nuclear structure of
odd-mass nuclei straddling $^{48}$Ca. Starting with the NN
interaction, that fits two-body scattering and bound state data we
evaluate the nuclear properties of $A = 47$ and $A = 49$ nuclei
while preserving all the underlying symmetries. Due to model space
limitations and the absence of 3-body interactions, we incorporate
phenomenological interaction terms determined by fits to $A = 48$
nuclei in a previous effort. Our modified Hamiltonian produces
reasonable spectra for these odd mass nuclei.  In addition to the
differences in single-particle basis states, the absence of a
single-particle Hamiltonian in our no-core approach complicates
comparisons with valence effective NN interactions. We focus on
purely off-diagonal two-body matrix elements since they are not
affected by ambiguities in the different roles for one-body
potentials and we compare selected sets of $fp$-shell matrix
elements of our initial and modified Hamiltonians in the harmonic
oscillator basis with those of a recent model $fp$-shell
interaction, the GXPF1 interaction of Honma, Otsuka, Brown and
Mizusaki. While some significant differences emerge from these
comparisons, there is an overall reasonably good correlation
between our off-diagonal matrix elements and those of GXPF1.
%\end{abstract}

\maketitle

\section{Introduction}
The low-lying levels of the $A=47 - 49$ nuclei have long been of
experimental and theoretical interest. On the one hand, extensive
experimental information about these nuclei is available
\cite{[BUR95]}-\cite{[ENDS]} and, on the other hand, this is a
suitable nuclear mass region for developing and testing effective
$fp$-shell Hamiltonians. Numerous detailed spectroscopic
calculations have been reported. For example, in Ref.
\cite{[MZPC96]}, using a shell model approach, Martinez-Pinedo,
Zuker, Poves and Caurier have performed full $fp$-shell
calculations for the $A=47$ and $A=49$ isotopes of Ca, Sc, Ti, V,
Cr and Mn.  They employed the KB3 interaction \cite{[KB3]} with
phenomenological adjustments and they performed complete
diagonalizations to obtain very good agreement with the
experimental level schemes, transition rates and static moments.
Extensive discussions of $fp$-shell effective Hamiltonians and
nuclear properties can be found in recent shell model review
articles \cite{[BAB01],[OtsukaHMSU01],[DJDean04],[Caurier05]}.

Our interest in these nuclei stems from our goal to extend the
no-core shell model (NCSM) applications to heavier systems than
previously investigated. Until recently, the NCSM, which treats
all nucleons on an equal footing, had been limited to nuclei up
through $A=16$. However, in a recent paper \cite{[VPSN06]} we
reported the first NCSM results for $^{48}$Ca, $^{48}$Sc and
$^{48}$Ti isotopes, with derived and phenomenological two-body
Hamiltonians. These three nuclei are involved in double-beta decay
of $^{48}$Ca, and the interest in developing nuclear structure
models for describing them is also related to the need for
accurate calculations of the nuclear matrix elements involved in
this decay. Our first goals were to see the limitations of such an
approach applied to heavier systems and how much improvement one
can obtain by adding phenomenological two-body terms involving all
nucleons. In brief, the results were the following
\cite{[VPSN06]}: i) one finds that the charge dependence of the
bulk binding energy of eight $A=48$ nuclei is reasonably described
with an effective Hamiltonian derived from CD-Bonn
interaction\cite{Machl} in the very limited $N_{\rm max}=0$ basis
space, while there is an overall underbinding by about $0.4$
MeV/nucleon; ii) the resulting spectra are too compressed compared
with experiment; iii) when isospin-dependent central terms plus a
tensor interaction are added to the Hamiltonian,  one achieves
accurate total binding energies for eight $A=48$ nuclei and
reasonable low-lying spectra for the three nuclei involved in
double-beta decay.  Only five input data were used to determine
the phenomenological terms - the total binding of $^{48}$Ca,
$^{48}$Sc, and $^{48}$Ti along with the lowest positive and
negative parity excitations of $^{48}$Ca.  The negative parity
calculations are performed in the $N_{\rm max}=1$ basis space.
Since the NCSM effective 2-body interaction is solely responsible
for the spectroscopy and involves the interactions of all $48$
nucleons, no single-particle energies are employed.

In the present paper we extend our previous approach to the
odd-$A$ isotopes $^{47}$Ca, $^{49}$Ca, $^{47}$Sc and $^{47}$K,
which differ by one nucleon from $^{48}$Ca. One of our goals is to
test whether the same modified effective 2-body Hamiltonian used
for $A=48$ isotopes, is able to describe these odd-$A$ nuclei and
whether experimentally known single-particle properties emerge in
a natural manner. A particular feature of the spectroscopy of
these odd nuclei is that the spin-orbit splitting gives rise to a
sizable energy gap in the $fp$-shell between the $f_{7/2}$ and
other orbitals ($p_{1/2}$, $p_{3/2}$, $f_{5/2}$) and we wanted to
see if this feature is reproduced in the NCSM. Also, in spite of
the differences in frameworks with and without a core, we aim to
compare selected aspects of our initial and modified Hamiltonian with a recent
$fp$-shell interaction, the GXPF1, developed by Honma, Otsuka,
Brown and Mizusaki \cite{[HOB04]}. We feel it is valuable to
compare various $fp$-shell interactions in order to understand
better their shortcomings and their regimes of applicability.
It is worth mentioning that our interaction and Honma et al. GXPF1
interactions were also tested recently within the framework of
spectral distribution theory in Ref. \cite{[SDV06a],[SDV06b]} to
illustrate their similarities and differences.

We note that direct comparisons of our matrix elements with those
of GXPF1 involve approximations that we have attempted to
minimize. We specifically list three aspects of the comparisons.
First, the $A$-dependence of GXPF1 is $A^{-0.3}$ while the
$A$-dependence of the derived effective interaction we employ is
not anticipated to be a simple scale factor.  Thus we focus our
attention on a narrow range of $A$ for the comparison.  Second, a
more precise comparison would involve the derivation of a pure
`valence-only' effective interaction and an appropriate scheme for
doing this has recently been shown to yield important 3-body
forces \cite{Lisetskiy08,Lisetskiy09}.  We hope the results
presented here help motivate that major undertaking.  Third, the
results we present may also be compared with the earlier
Brueckner-based matrix elements \cite{Hjorth-Jensen95} for the
same region since our phenomenological adjustments may be expected
to simulate the physics accounted for in the perturbative
treatment of core-polarization.

Our paper is organized as follows: in Section 2 we give a short review of the NCSM
approach and we refer the reader to the bibliography for more details.
Section 3 is devoted to the presentation of our results. Binding energies, excitation spectra,
single-particle characteristics, monopole matrix elements and matrix element correlations
are discussed in subsections along with corresponding figures. In the
last Section we present the conclusion and the outlook of our work.

\section{NO-CORE SHELL MODEL}

The NCSM~\cite{ZBV,NB96,NB98,NB99,NCSM12,NCSM6} is
based on an effective Hamiltonian derived from realistic
``bare'' interactions and acting within a finite Hilbert space.
All $A$-nucleons are treated on an
equal footing.  The approach is both computationally
tractable and demonstrably convergent to the exact result
of the full (infinite) Hilbert space.

Initial investigations used two-body interactions~\cite{ZBV} based
on a G-matrix approach. Later, a similarity transformation
procedure based on Okubo's pioneering work~\cite{Vefftot} was
implemented to derive two-body and three-body effective
interactions from realistic $NN$ and $NNN$ interactions.
Diagonalization and evaluation of observables from effective
operators created with the same transformations are carried out on
high-performance parallel computers.

\subsection{Effective Hamiltonian}

In order to clarify the distinctions from the conventional shell
model with a core, we briefly outline the {\it ab initio} NCSM
approach with $NN$ interactions alone and point the reader to the
literature for the extensions to include $NNN$ interactions. We
begin with the purely intrinsic Hamiltonian for the $A$-nucleon
system, i.e.,
\begin{equation}\label{ham}
H_A= T_{\rm rel} + {\cal V} =
\frac{1}{A}\sum_{i<j}^A \frac{(\vec{p}_i-\vec{p}_j)^2}{2m}
+ \sum_{i<j=1}^A V_{\rm N}(\vec{r}_i-\vec{r}_j) \; ,
\end{equation}
where $m$ is the nucleon mass and $V_{\rm
N}(\vec{r}_i-\vec{r}_j)$, the $NN$ interaction, with both strong
and electromagnetic components. Note the absence of a
phenomenological single-particle potential. We may use either
coordinate-space $NN$ potentials, such as the Argonne potentials
\cite{GFMC} or momentum-space dependent $NN$ potentials, such as
the CD-Bonn \cite{Machl}.

Next, we add to (\ref{ham}) the center-of-mass Harmonic Oscillator (HO) Hamiltonian
$H_{\rm CM}=T_{\rm CM}+ U_{\rm CM}$, where
$U_{\rm CM}=\frac{1}{2}Am\Omega^2 \vec{R}^2$,
$\vec{R}=\frac{1}{A}\sum_{i=1}^{A}\vec{r}_i$.
At convergence, the added  $H_{\rm CM}$ term
has no influence on the intrinsic properties.
However, when we introduce the cluster approximation below,
the added  $H_{\rm CM}$ term
facilitates convergence to exact results
with increasing basis size.
The modified Hamiltonian, with pseudo-dependence on the HO
frequency $\Omega$, can be cast as:
\begin{equation}
%\begin{multline}
\label{hamomega}
H_A^\Omega= H_A + H_{\rm CM}=\sum_{i=1}^A \left[ \frac{\vec{p}_i^2}{2m}
+\frac{1}{2}m\Omega^2 \vec{r}^2_i
\right]\\
%+ \sum_{i<j=1}^A \left[ V_{\rm N}(\vec{r}_i-\vec{r}_j)
+ \sum_{i<j=1}^A \left[ V_{\rm N}(ij)
-\frac{m\Omega^2}{2A}
(\vec{r}_i-\vec{r}_j)^2
\right] \; .
%\end{multline}
\end{equation}

Next, we introduce a unitary transformation, which is designed to
accommodate the short-range two-body correlations in a nucleus, by
choosing an anti-hermitian operator $S$, acting only on intrinsic
coordinates, such that ${\cal H} = e^{-S} H_A^\Omega e^{S}$.
%
%\begin{equation}\label{UMOAtrans}
%{\cal H} = e^{-S} H_A^\Omega e^{S} \; .
%\end{equation}
%
In this approach, $S$ is determined by the requirements that
${\cal H}$ and $H_A^\Omega$ have the same symmetries and
eigenspectra over the subspace ${\cal K}$ of the full Hilbert
space. In general, both $S$ and the transformed Hamiltonian are
$A$-body operators. The simplest, non-trivial approximation to
${\cal H}$ is to develop a two-body $(a=2)$ effective Hamiltonian,
where the upper bound of the summations ``$A$'' is replaced by
``$a$'', but the coefficients remain unchanged.
%The next improvement is to develop a three-body effective Hamiltonian, $(a=3)$.

%We then have an approximation at a fixed level of clustering, $a$,
%with $a \leq A$.
%
%\begin{equation}\label{UMOAexpan}
%{\cal H} = {\cal H}^{(1)} + {\cal H}^{(a)} =  \sum_{i=1}^{A} h_i +
%\frac{{A \choose 2}}{{A \choose a}{a \choose 2}}
%\sum_{i_{1}<i_{2}< \ldots <i_{a}}^{A}\tilde{V}_{i_{1}i_{2} \ldots i_{a}} \; ,
%\end{equation}
%with
%
%\begin{equation}\label{UMOAexplterms}
%\tilde{V}_{12 \ldots a} = e^{-S^{(a)}}H^{\Omega}_{a}e^{S^{(a)}}
%- \sum_{i=1}^a h_i \; , \\
%\end{equation}
%
%and $S^{(a)}$ is an $a$-body operator; $H^{\Omega}_{a} = h_1+h_2+h_3+ \ldots
%+h_{a}+V_{a}$,
%and $V_{a} = \sum_{i<j}^{a} V_{ij}$.
%There is no sum over ``$a$'' in Eq. (\ref{UMOAexpan}).

%For the many-body basis, we adopt Slater determinants that are
%eigenstates of the sum of one-body Hamiltonians $\sum_{i=1}^A
%h_i$. These one-body Hamiltonians  are also taken to be harmonic
%oscillators.

The full Hilbert space is divided into a finite model space
(``$P$-space'') and a complementary infinite space (``$Q$-space''), using
the projectors $P$ and $Q$ with $P+Q=1$. We determine the transformation
operator $S_{a}$ from the decoupling condition
%
%\begin{equation}\label{UMOAdecoupl}
$Q_{a} e^{-S^{(a)}}H^{\Omega}_{a}e^{S^{(a)}} P_{a} = 0$
%   \; ,
%\end{equation}
%
and the simultaneous restrictions $P_a S^{(a)} P_a = Q_a S^{(a)} Q_a =0$.
The $a$-nucleon-state projectors ($P_a, Q_a$)
%appear in Eq. (\ref{UMOAdecoupl}). Their definitions
follow from the
definitions of the $A$-nucleon projectors $P$, $Q$.

In the limit $a \rightarrow A$, we obtain the exact solutions for
$d_P$ states of the full problem for any finite basis space, with
flexibility for choice of physical states subject to certain
conditions~\cite{Viaz01}. This approach has a significant residual
freedom through an arbitrary residual $P_a$--space unitary
transformation that leaves the $a$-cluster properties invariant.
Of course, the $A$-body results obtained with the $a$-body cluster
approximation are not invariant under this residual
transformation.  It may be worthwhile, in a future effort, to
exploit this residual freedom to accelerate convergence in
practical applications.

The model space, $P_2$, is defined by $N_{\rm m}$ via the maximal
number of allowed HO quanta of the $A$-nucleon basis states, $N_{\rm
M}$, where the sum of the nucleons'
$2n+l\leq N_{\rm m}+N_{\rm spsmin}=N_{\rm M}$, and where $N_{\rm
spsmin}$ denotes the minimal possible HO quanta of the spectators,
nucleons not involved in the interaction.
%For example, $^{10}$B,
For example,  in $^{10}$B we have
$N_{\rm spsmin}=4$ as there are 6 nucleons in the $0p$-shell
in the lowest HO configuration and
%, e.g.,
$N_{\rm m}=2+N_{\rm max}$, where $N_{\rm max}$
represents the maximum HO quanta of the many-body excitation
above the unperturbed ground-state configuration.
For $^{10}$B
%,
with
 $N_{\rm M}=12, N_{\rm m}=8$
for an $N_{\rm max}=6$ or ``$6\hbar\Omega$'' calculation.
With the cluster approximation,
a dependence of our results on $N_{\rm max}$
(or equivalently, on $N_{\rm m}$ or $N_{\rm M}$) and on
$\Omega$ arises.
%For a fixed cluster size $a$,
%the smaller the basis space, the larger the
%dependence on $\Omega$.
The residual $N_{\rm max}$ and $\Omega$ dependences
will infer the uncertainty in our results
arising from effects associated with increasing $a$ and/or
effects with increasing $N_{\rm max}$.  In the present work, we
retain the $N_{\rm max}=0$ basis space and $\hbar\Omega=10 MeV$
employed in Ref. \cite{[VPSN06]}.

At this stage
we also add the term $H_{CM}$ again with a large positive coefficient
(the Lagrange multiplier)
to separate the physically interesting states with $0s$ CM motion from
those with excited CM motion. We diagonalize the effective Hamiltonian
with the m-scheme Lanczos method to obtain the $P$-space
eigenvalues and eigenvectors \cite{Vary_code}.
%We retain only the states with pure
%$0s$ CM motion when evaluating observables.
All observables are then evaluated free of CM motion effects
\cite{Vary_code}. In principle, all observables require the same
transformation as implemented for the Hamiltonian.  We obtain
small renormalization effects on long range operators such as the
rms radius operator and the $B(E2)$ operator when we transform
them to $P$-space effective operators at the $a=2$ cluster
level~\cite{NCSM12,Stetcu}. On the other hand, when $a=2$,
substantial renormalization was observed for the kinetic energy
operator \cite{benchmark01}. and for higher momentum transfer
observables \cite{Stetcu}.

Recent applications and extensions include:
\begin{itemize}
\item[(a)] spectra and transition rates in $p$-shell nuclei \cite{Nogga06};
\item[(b)] comparisons between NCSM and Hartree-Fock \cite{Hasan03};
%\item[(c)] di-neutron correlations in the $^6$He halo
%   nucleus \cite{Atramentov05};
\item[(c)] neutrino cross sections on $^{12}$C \cite{HayesPRL};
\item[(d)] novel $NN$ interactions using inverse scattering theory
plus NCSM\cite{Shirokov04, Shirokov05,Shirokov06}; \item [(e)]
solving for light nuclei with chiral effective interactions
\cite{Navratil07} ; \item[(f)] studies of alternative converging
sequences \cite{Forssen08}; \item[(g)] solving for light nuclei
with alternative renormalization schemes \cite{Bogner08};
\item[(h)] development of effective interactions
\cite{Lisetskiy08} and operators  \cite{Lisetskiy09} with a core;
\item[(i)] solving for light nuclei with bare $NN$ interactions
with extrapolation to the infinite matrix limit \cite{Maris09};
\item[(j)] phenomenology of hadronic structure and quantum field
theory \cite{Vary06}.
\end{itemize}

We close this theory overview by referring to the added
phenomenological $NN$ interaction terms found adequate for
obtaining good descriptions of $A=48$ nuclei \cite{[VPSN06]}.
Three terms are added - central modified Gaussians with isospin
dependent strengths and a tensor term.  The forms were chosen
primarily for their simplicity.  However, based on experience in
light nuclei, it is desirable to include a tensor term in order to better
fit the spin-sensitive properties.
In the hope of obtaining a
simple NCSM Hamiltonian for the BE and spectra of $A = 47$, $48$
and $49$ nuclei, we assume the added NN interaction terms are
$A$-independent.  That is, they are taken to complement
the $A$-dependent NCSM $H_{eff}$ in a very simple manner.
NCSM results obtained below
with the modified Hamiltonian are referred to with ``CD-Bonn + 3
terms'' results. It is our hope that these terms accommodate, to a
large extent, the missing many-body forces, both real and
effective. This hope will be tested in the future when increasing
computational resources will allow larger basis spaces, improved
$a=3$ and $a=4$ calculations as well as the introduction of true
$NNN$ and $NNNN$ potentials.

\section{RESULTS AND DISCUSSION}

\subsection{Binding energies}

\begin{figure}[htb]
\begin{minipage}[t]{80mm}
\framebox[79mm]{\rule[0mm]{0mm}{52mm}\includegraphics[scale=0.32]{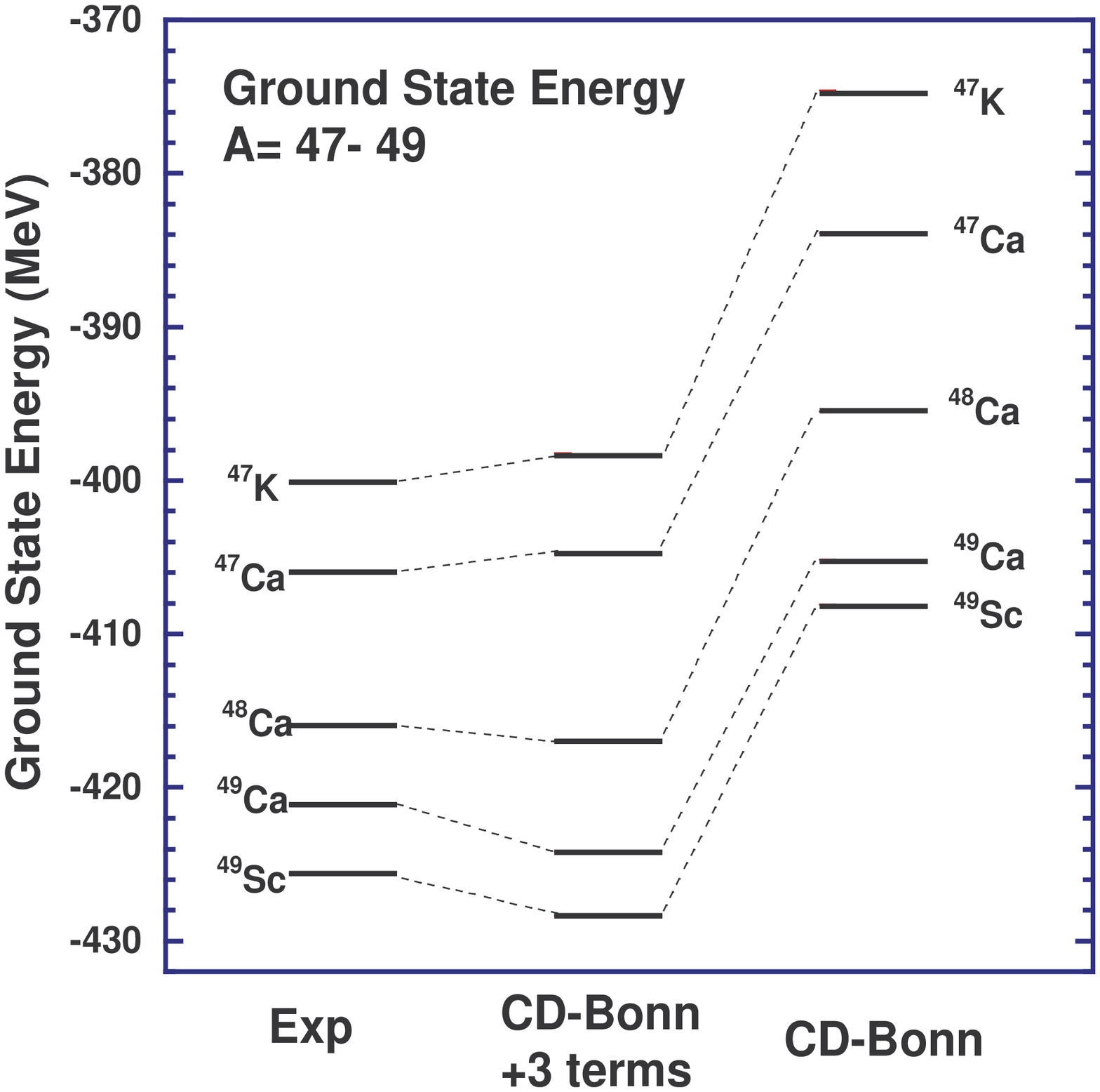}}
\caption{The experimental and theoretical ground state energy
levels for $A=47-49$. The results in the second and third columns
are labelled by their Hamiltonians.}
\label{fig:Ground-state-energy}
\end{minipage}
\hspace{\fill}
\begin{minipage}[t]{84mm}
\framebox[80mm]{\rule[0mm]{0mm}{52mm}\includegraphics[scale=0.38]
%%{Ca49.eps}}
%%{Ca49_trial.eps}}
{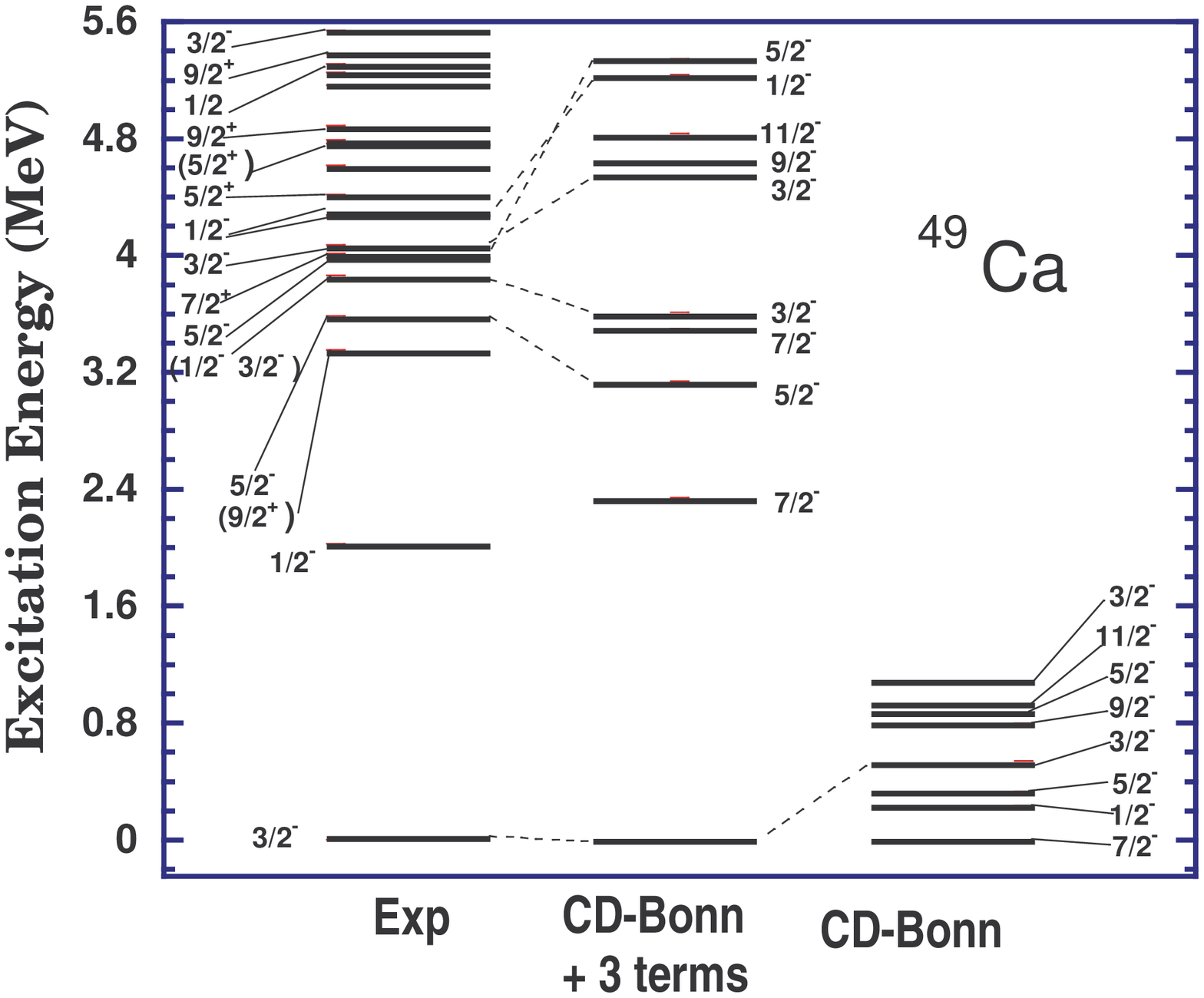}}
\caption{Experimental and theoretical excitation energy levels for
$^{49}{Ca}$. Both CD-Bonn and CD-Bonn+3terms results are presented.}
\label{fig:Ca49}
\end{minipage}
\end{figure}

First we present the calculated total interaction energies
(Hamiltonian ground state eigenvalues) in Fig.
\ref{fig:Ground-state-energy} which we compare with experiment.
One observes that ground states calculated with our derived {\it
ab initio} $H_{eff}$ lie above the experimental values by
approximately $20$ MeV. This shift is similar to that observed in
the case of all $A=48$ isotopes \cite{[VPSN06]}. We note that with
CD-Bonn we have nearly the same increase in binding from $^{47}$Ca
to $^{48}$Ca as from $^{48}$Ca to $^{49}$Ca, which signals a lack
of  subshell closure.

For the modified Hamiltonian (CD-Bonn + 3 terms) the NCSM produces
reasonable agreement with experiment with deviations much less
than 1\% as seen in Fig. \ref{fig:Ground-state-energy}.  There is
a simple spreading of the theoretical ground states relative to
experiment. In particular,  we now observe the desired subshell
closure condition where the increased binding from $^{47}$Ca to
$^{48}$Ca significantly exceeds that from  $^{48}$Ca to $^{49}$Ca.

%Figures 1-2

%Figures 3-4
\begin{figure}[htb]
\begin{minipage}[t]{75mm}
\framebox[74mm]{\rule[0mm]{0mm}{52mm}\includegraphics[scale=0.32]
{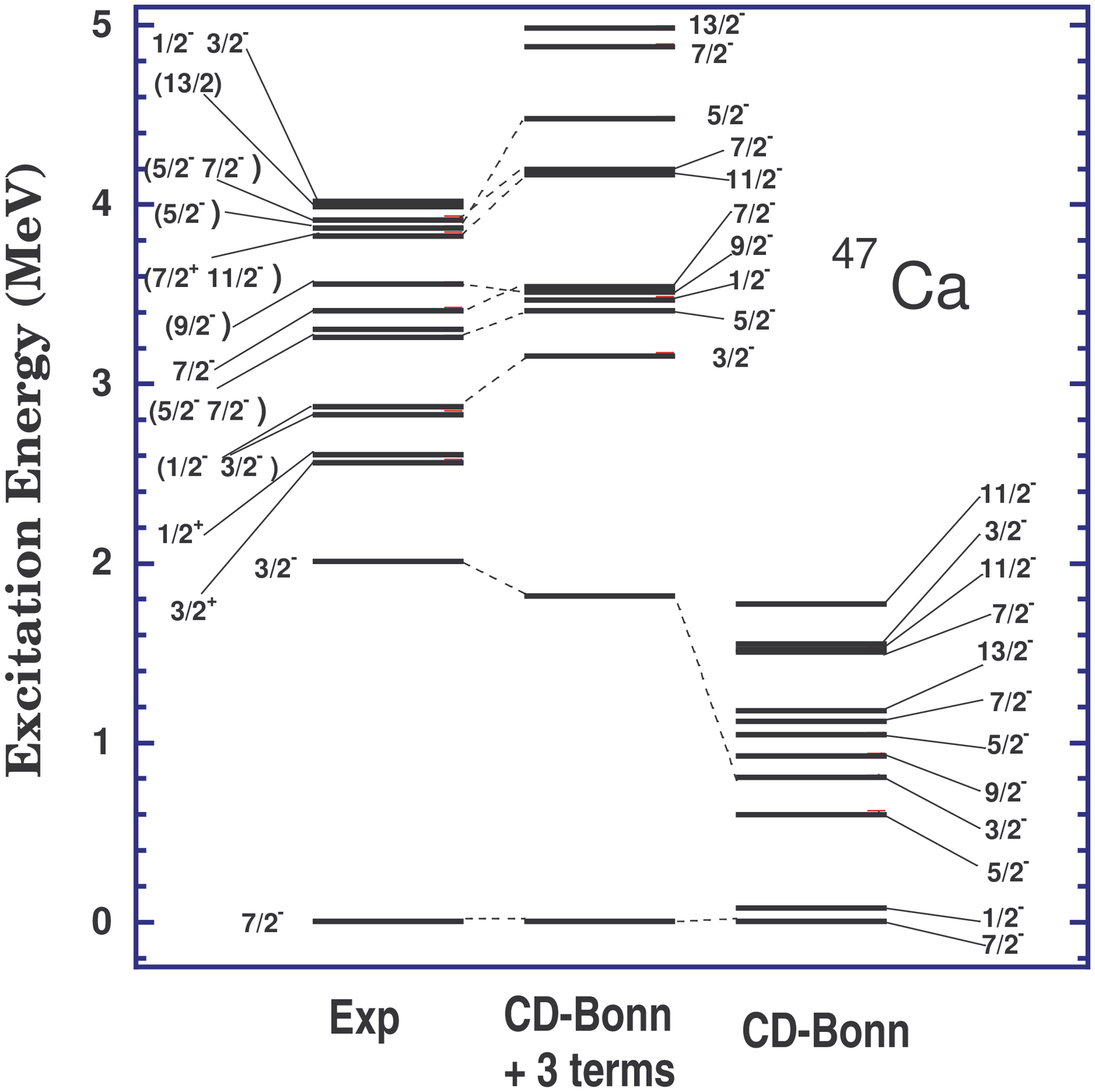}}
\caption{Experimental and theoretical excitation energy levels for
$^{47}{Ca}$. Both CD-Bonn and CD-Bonn+3terms results are presented.}
\label{fig:Ca47}
\end{minipage}
\hspace{\fill}
\begin{minipage}[t]{75mm}
\framebox[79mm]{\rule[0mm]{0mm}{52mm}\includegraphics[scale=0.32]
{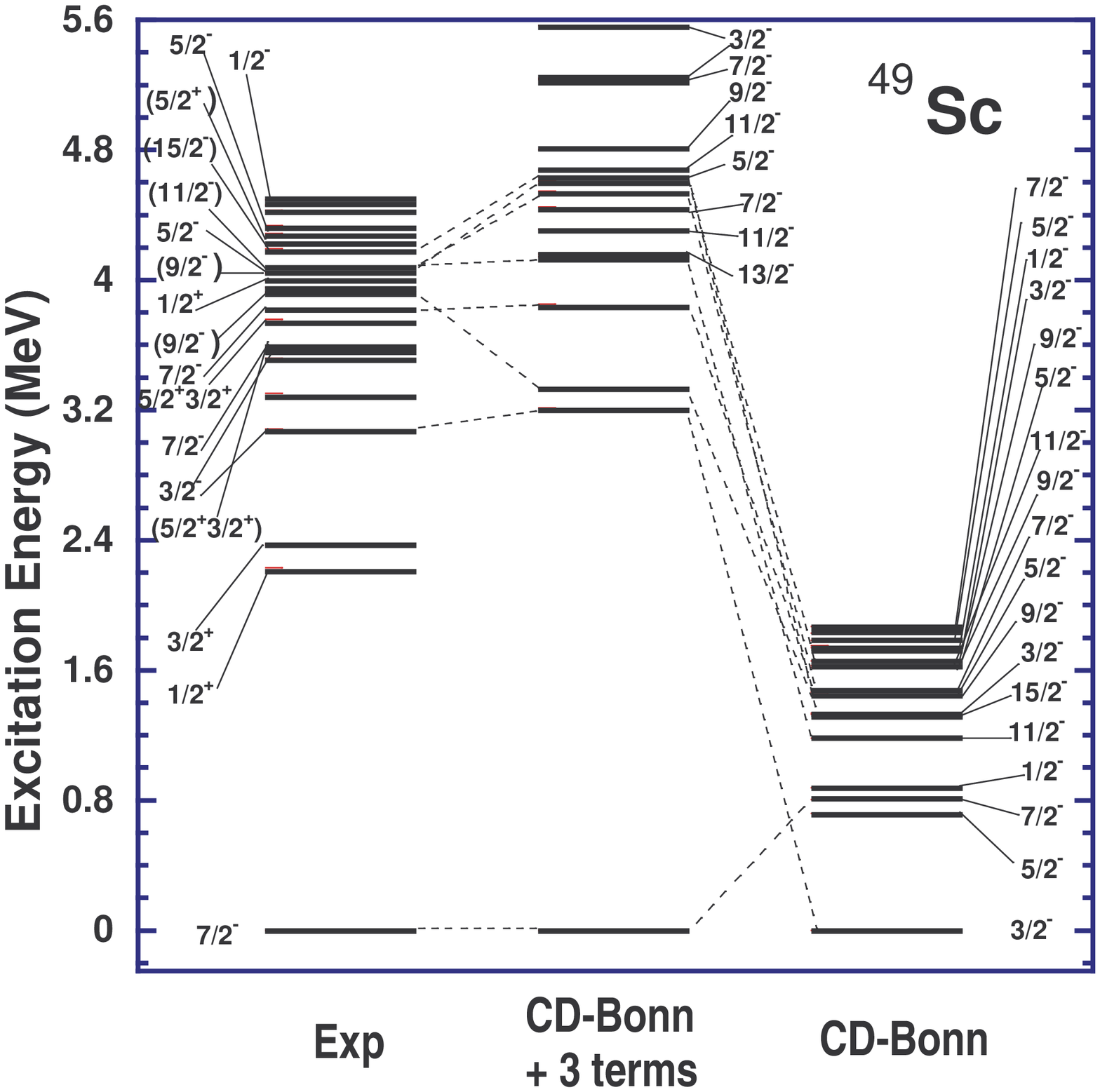}}
\caption{Experimental and theoretical excitation energy levels for
$^{49}{Sc}$. Both CD-Bonn and CD-Bonn+3terms results are presented.}
\label{fig:Sc49}
\end{minipage}
\end{figure}

%Figures 5-6
\begin{figure}[htb]
\begin{minipage}[t]{75mm}
\framebox[74mm]{\rule[0mm]{0mm}{52mm}\includegraphics[scale=0.32]
{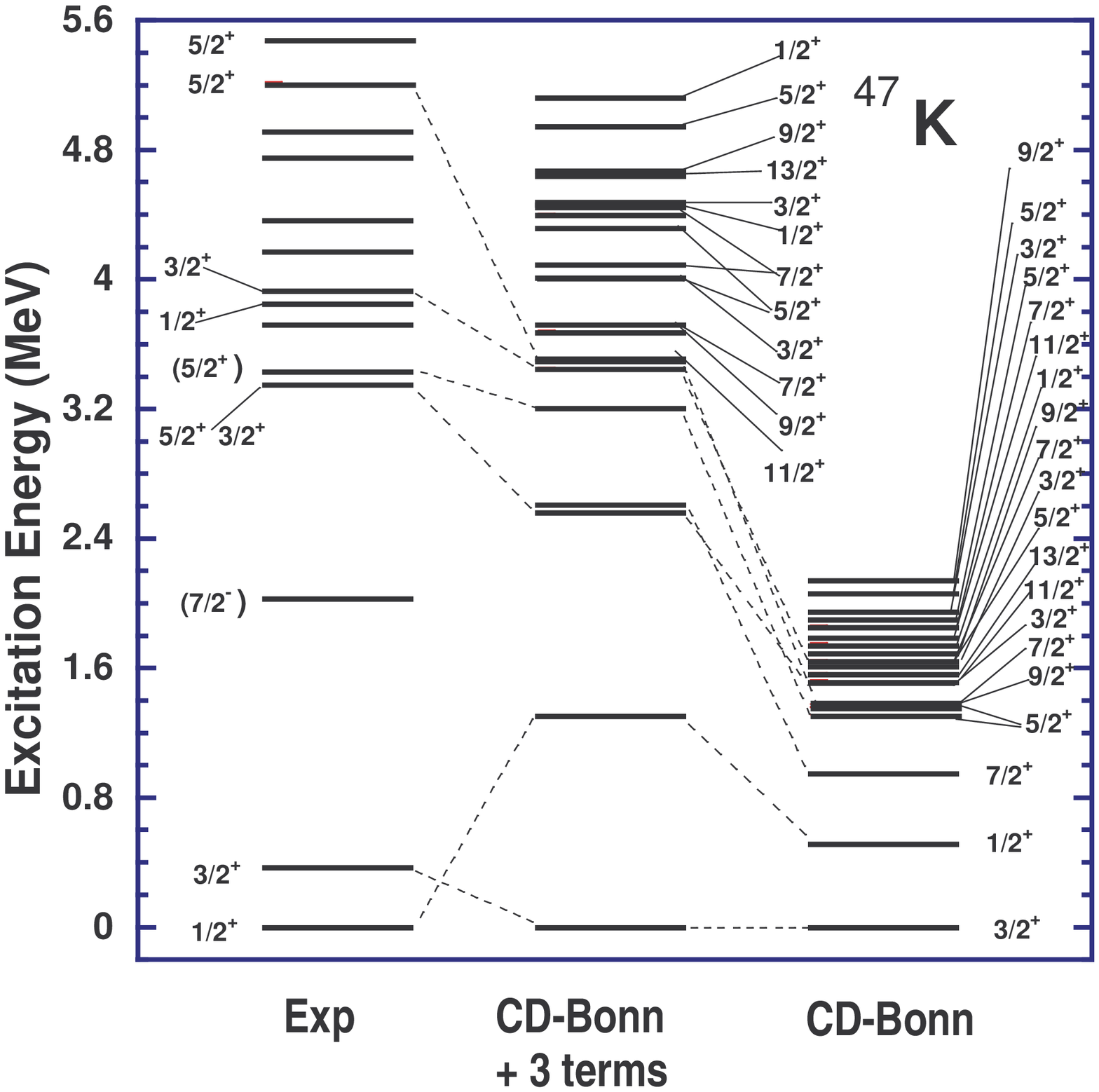}}
\caption{Experimental and theoretical excitation energy levels for
$^{47}{K}$. Both CD-Bonn and CD-Bonn+3terms results are presented.}
\label{fig:K47}
\end{minipage}
\hspace{\fill}
%
% This must not be the final figure as there is no dashed line for the Fermi
% energy.
\begin{minipage}[t]{75mm}
\framebox[74mm]{\rule[0mm]{0mm}{52mm}\includegraphics[scale=0.32]
{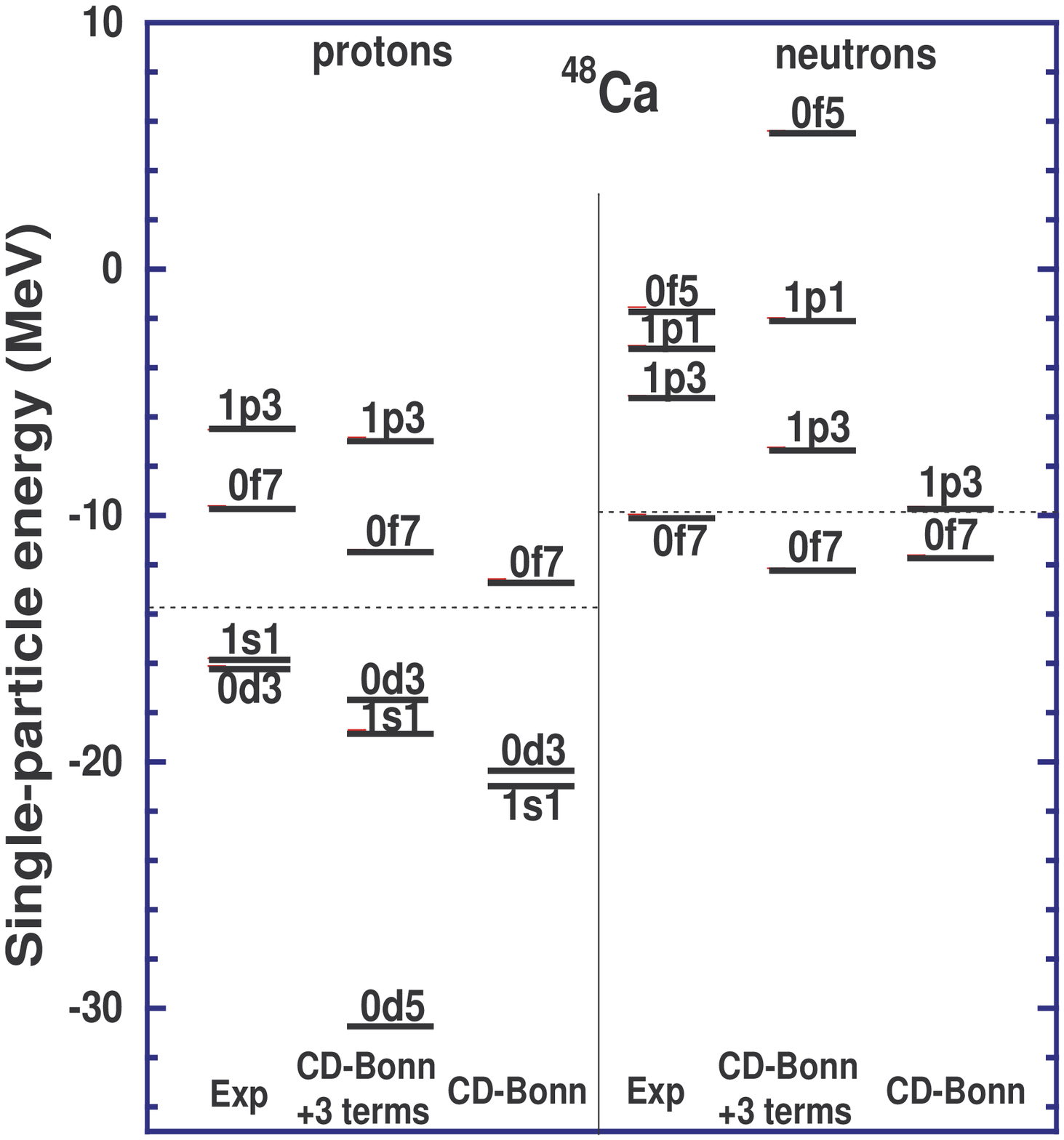}}
\caption{Experimental and theoretical levels that are dominantly
single-proton and single-neutron particles or holes of $^{48}{Ca}$.
The levels are labeled by  (n, l, 2j), and the dashed lines are the
Fermi energies.}
\label{fig:Single-particle-energy}
\end{minipage}
\end{figure}

%Removed Fig. 7-10
%Figures CDB3_vs_CDB 11-12
% Fig. 7
\begin{figure}[htb]
\begin{minipage}[t]{80mm}
\framebox[74mm]{\rule[0mm]{0mm}{52mm}\includegraphics[scale=0.43]
{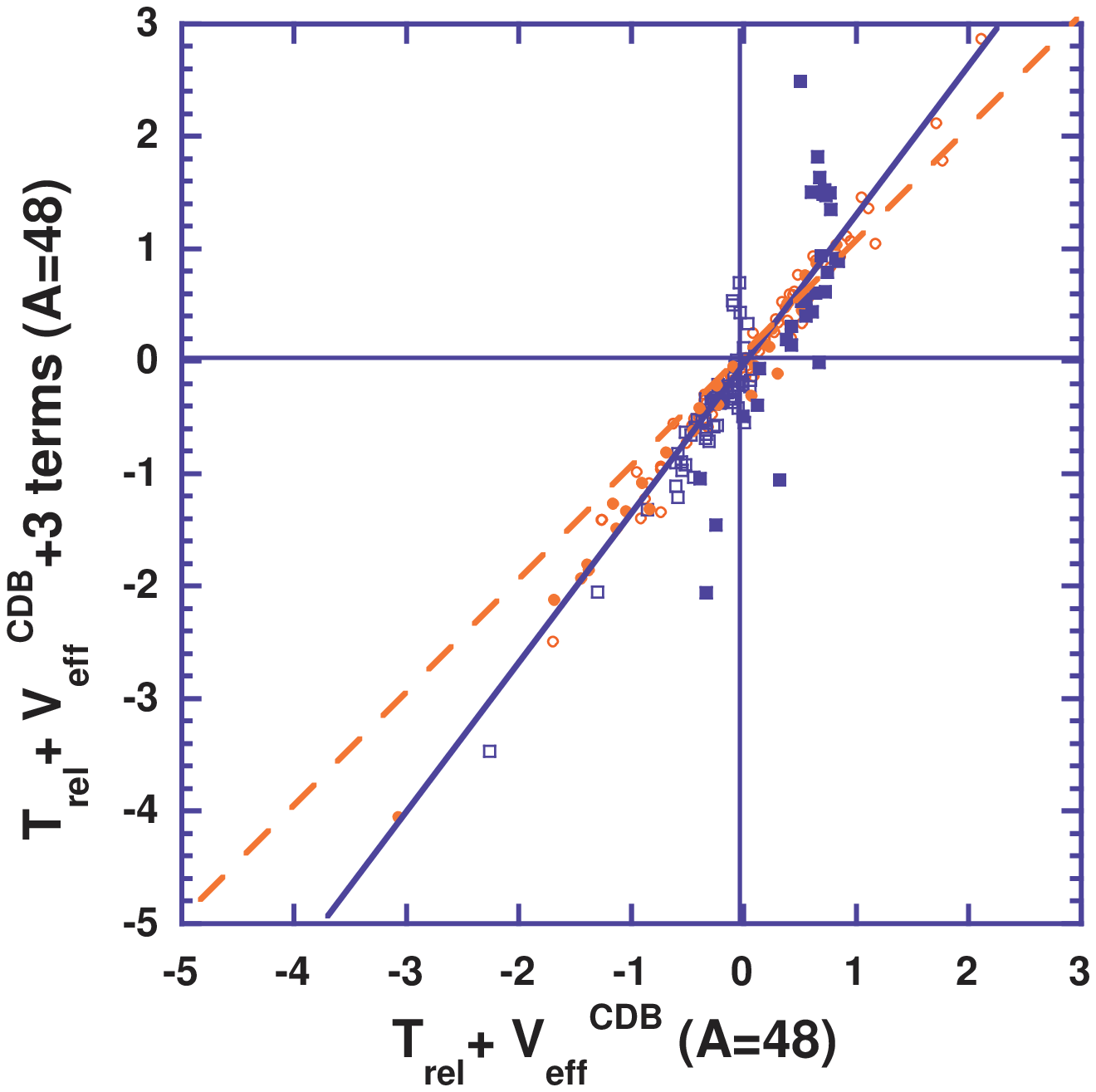}}
\caption{(color online) Correlation of $V(abcd; JT)$ matrix
elements between CD-Bonn+3terms and CD-Bonn. The matrix elements
of $T=0$ and $T=1$ are shown by circles and squares, respectively.
The filled circles and the filled squares are for all $V(abab;
JT)$ matrix elements that contribute to the monopole $V(ab; T)$.
The open circles and open squares are for all the remaining matrix
elements. There are no monopole shifts. The solid straight line
represents a linear fit to all the matrix elements. The diagonal
dashed line represents the reference correlation line at
$45$-degrees.} \label{CDB3CDBallME}
\end{minipage}
\hspace{\fill}
\begin{minipage}[t]{75mm}
\framebox[74mm]{\rule[0mm]{0mm}{52mm}\includegraphics[scale=0.43]
%%{Fig.14_CDB_3_CDB_after_eliminate_all_abab_ME.eps}}
{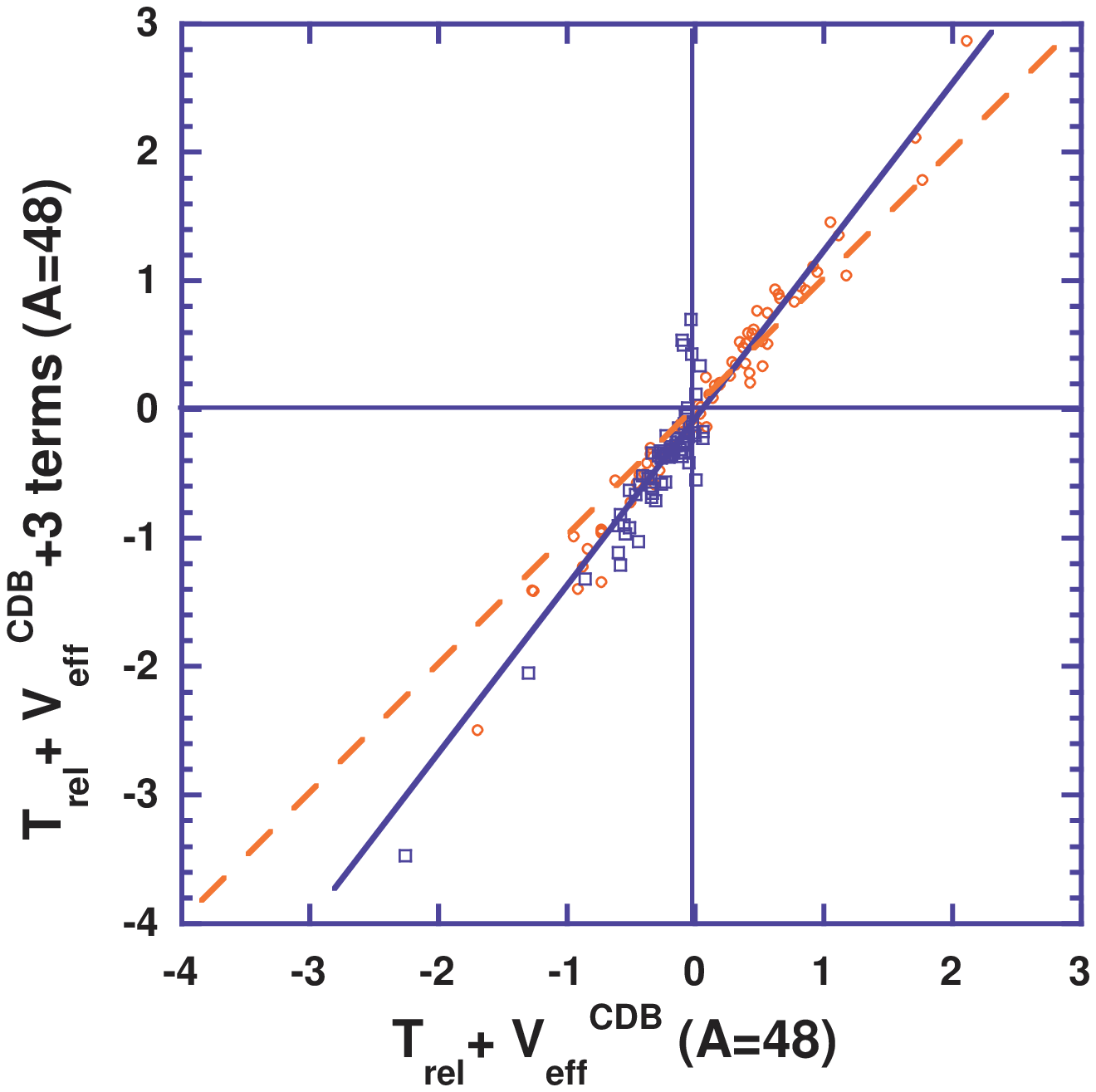}}
\caption{(color online) Correlation of $V(abcd; JT)$ between
CD-Bonn+3terms and CD-Bonn, where we removed all $V(abab; JT)$
matrix elements that contribute to the monopole $V(ab; T)$. The
solid straight line represents a linear fit to all the plotted
points. The open circles and the open squares stand for $T=0$ and
$T=1$, respectively. The diagonal dashed line represents the
reference correlation line at $45$-degrees.}
\label{CDB3vsCDBafterelimababME}
\end{minipage}
\end{figure}

%Figures CDB_vs_G 13-14
\begin{figure}[htb]
\begin{minipage}[t]{80mm}
\framebox[74mm]{\rule[0mm]{0mm}{52mm}\includegraphics[scale=0.43]
{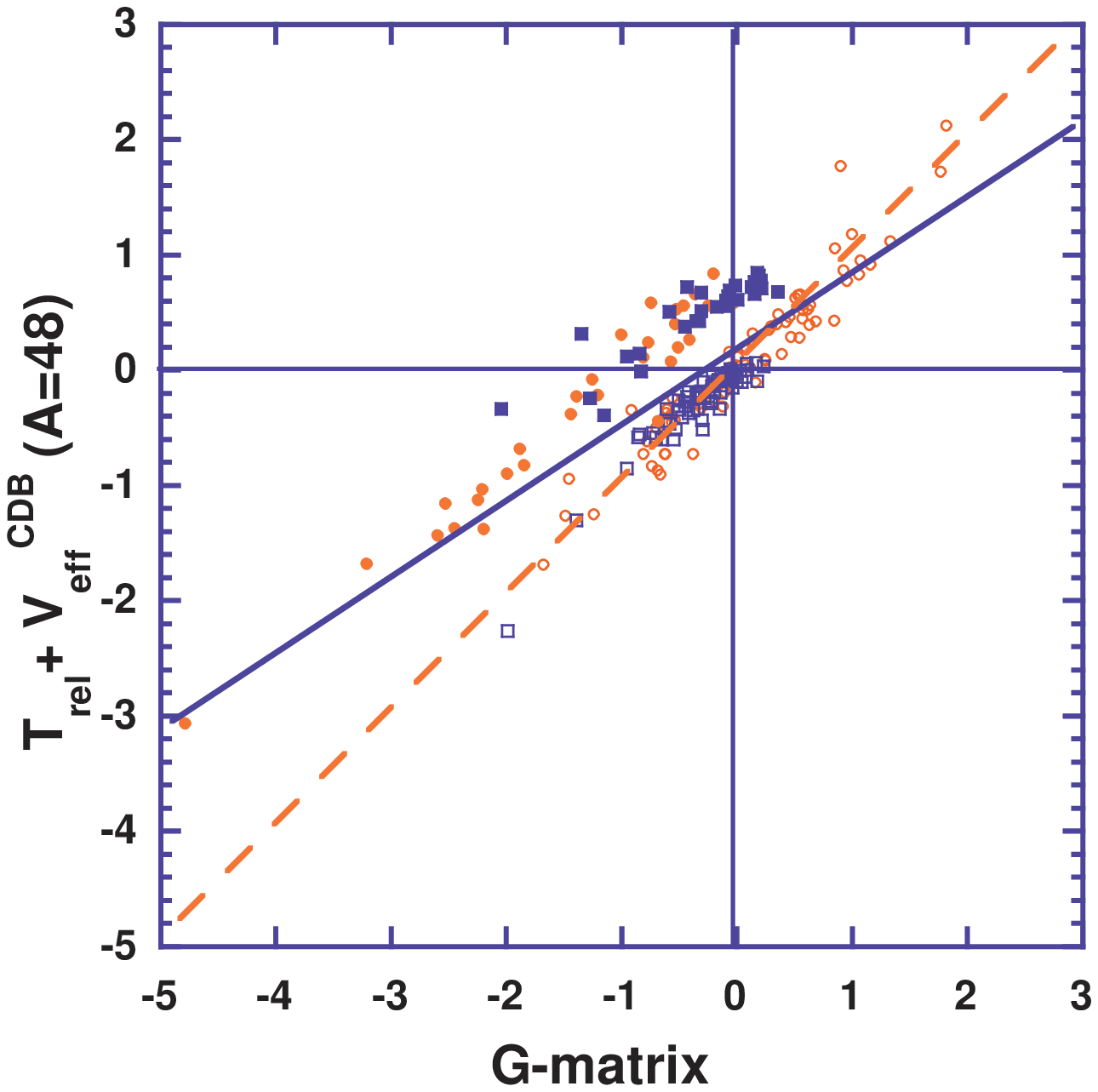}}
\caption{(color online) Correlation of $V(abcd; JT)$ matrix
elements between CD-Bonn and G-matrix. See the caption to Fig.
\ref{CDB3CDBallME}.} \label{CDBvsGallME}
\end{minipage}
\hspace{\fill}
\begin{minipage}[t]{75mm}
\framebox[74mm]{\rule[0mm]{0mm}{52mm}\includegraphics[scale=0.43]
{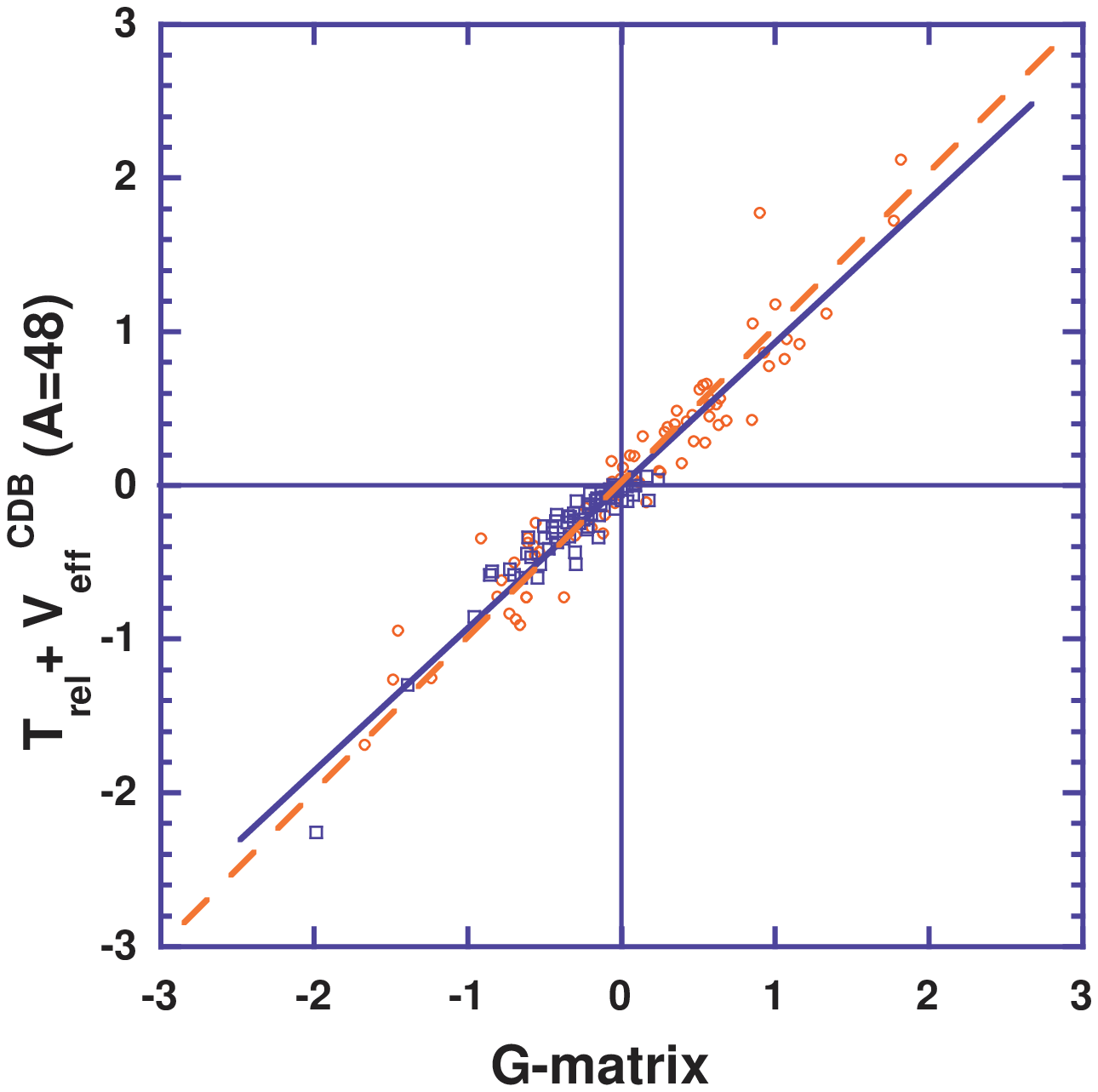}}
\caption{(color online) Correlation of $V(abcd; JT)$ between
CD-Bonn and G. See the caption to Fig.
\ref{CDB3vsCDBafterelimababME}.} \label{CDBvsGafterelimababME}
\end{minipage}
\end{figure}

%Figure 15 removed
%Figure 16
\begin{figure}[htb]
\begin{minipage}[t]{75mm}
\framebox[74mm]{\rule[0mm]{0mm}{52mm}
\includegraphics[scale=0.43]
%{Exclusive_off_diagonal_last_version.eps}}
{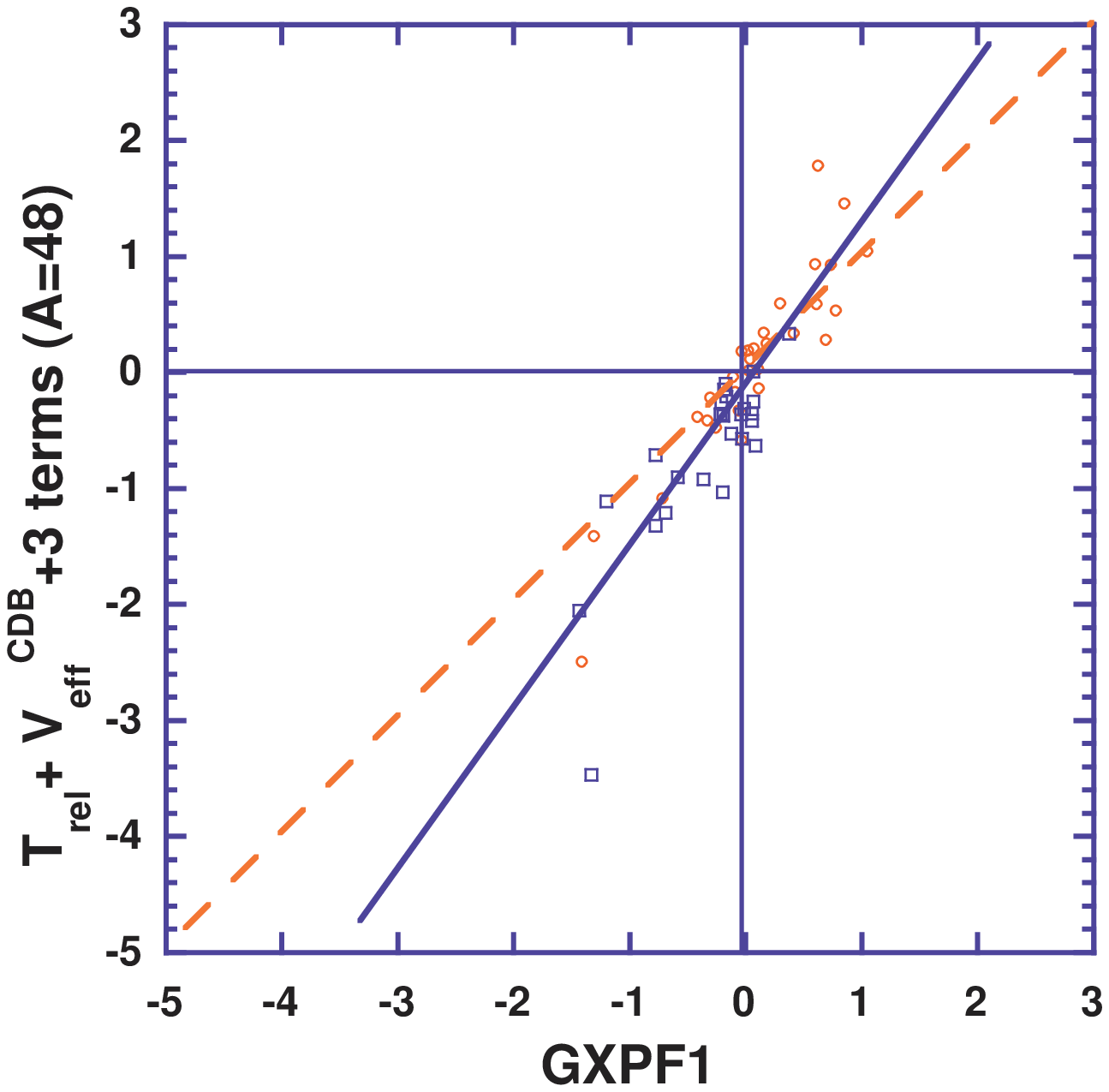}} \caption{(color
online) Correlation of the matrix elements $V(abcd; JT)$ between
CD-Bonn+3terms and GXPF1, where we retain only the off-diagonal
matrix elements, i.e. the $56$ matrix elements that cannot
contribute to a single-particle Hamiltonian (see text). The open
circles and the open squares stand for $T=0$ and $T=1$,
respectively. The thick solid line represents a linear fit to
all the plotted points. The diagonal dashed line represents the
reference correlation line at $45$-degrees.}
\label{CDB3GXPF1TrueOff-Diag}
\end{minipage}
\end{figure}

%Figure 17&18 removed

\subsection{Excitation energy spectra}

\noindent The excitation energy spectra for $^{49}$Ca, $^{47}$Ca,
$^{49}Sc$ and $^{47}$K are shown in Figs.
\ref{fig:Ca49}-\ref{fig:K47} respectively. In every case the {\it
ab initio} NCSM results with CD-Bonn are far too compressed
relative to experiment - a feature also seen in the $A=48$ results
\cite{[VPSN06]}.  Here, we trace this primary defect to the
inferred properties of the neutron orbits. That is, the incorrect
ground state spin seen in Fig. \ref{fig:Ca49} and the absence of a
significant excitation energy gap in Fig. \ref{fig:Ca47} indicate
the spin-orbit splitting of the neutrons is insufficient to
provide proper subshell closure at the neutron $0f_{7/2}$ orbit.
This defect is rectified in the results with CD-Bonn + 3 terms as
seen by the corresponding spectra in Figs. \ref{fig:Ca49} and
\ref{fig:Ca47}. Similar tendencies have been seen before with
valence G-matrix interactions and identified as a problem with the
$L^{2}$ dependence of the single-particle states
\cite{[MZPC96],[Caurier05]}.

The CD-Bonn results in Figs. \ref{fig:Sc49} and \ref{fig:K47} are
more difficult to interpret due to the glaring deficiencies just
mentioned for the neutrons with the CD-Bonn Hamiltonian.  We will
show below that the proton shell closure is better established
with CD-Bonn. This supports the assertion that the main
deficiencies seen in the third columns of Figs. \ref{fig:Sc49} and
\ref{fig:K47} are indeed likely to reside with the inferred
neutron spin-orbit splitting problem.

The modified Hamiltonian provides greatly improved spectra for all
four nuclei as seen in the second columns of Figs.
\ref{fig:Ca49}-\ref{fig:K47}. It is to be noted that these nuclei
were not involved in the fitting procedure used to determine the
parameters of the added phenomenological terms.  Perhaps the most
significant remaining deficiency is the incorrect ground state
spin for $^{47}$K as seen in Fig. \ref{fig:K47}. This is the first
case of a nucleus in the region of $A=47$ to $A=49$ (12 nuclei
studied to date) where we did not obtain the correct ground state
spin with CD-Bonn + 3 terms Hamiltonian.

\subsection{Single-particle characteristics}

In order to better understand the underlying physics of
our NCSM results, we investigate the single-particle-like
properties of our solutions.

In a simple closed-shell nucleus, we expect the leading
configuration of the ground state solution in our m-scheme
treatment to be a single Slater determinant.  Single-particle (or
hole) excitations should be easily identified by the character of
their leading configurations - i.e. a single-particle creation (or
destruction) operator acting on the ground state Slater
determinant of the reference nucleus $^{48}$Ca. For our odd-mass
nuclei, this is the character we seek.  That is, we take the
standard phenomenological shell model configuration of a single
Slater determinant with a closed $sd$-shell for the protons and a
closed $f_{7/2}$ subshell for the neutrons and look for the
appropriate states which have a single nucleon added to (or
subtracted from) that Slater determinant.  We accept states as
``single-particle-like'' when we find one with a leading
configuration having more than 50\% probability to be in the
simple configuration just described.  When the majority weight is
distributed over a few states, we use the centroid and we discuss
those cases in some detail below.  We were not successful  in
locating all the expected single-particle-like and
single-hole-like states. That is, those absent from our
presentation below were spread among a large number of
eigenstates.

For a closed-shell nucleus $(Z,N)$ the single-particle energies
(SPE) for states above the Fermi surface are related to the
binding energy
differences:\\

$e_p^{>} = BE(Z,N)-BE^{*}(Z+1,N)$\label{equationa}\\
and

$e_n^{>} = BE(Z,N)-BE^{*}(Z,N+1)$.\label{equationb}\\

The SPE for sates below the Fermi surface are given by

$e_p^{<} = BE^{*}(Z-1,N)-BE(Z,N)$\label{equationc}\\
and

$e_n^{<} = BE^{*}(Z,N-1)-BE(Z,N)$.\label{equationd}\\

The BE are ground state binding energies which are taken as
positive values, and $e$ will be negative for bound states.
$(BE^{*} = BE-E_{x})$ is the ground state binding energy minus the
excitation energy of the excited states associated with the
single-particle states.

Experimental SPE's and the results of our analysis are shown in
Fig. \ref{fig:Single-particle-energy}. The experimental SPE's for
protons and neutrons follow B.A.Brown's analysis \cite{[BAB01]}.
To guide the eye, we draw a horizontal line to indicate the
vicinity of the Fermi surfaces for the protons and neutrons.

Fig. \ref{fig:Single-particle-energy} shows that proton shell
closure is established with both Hamiltonians, the CD-Bonn and the
CD-Bonn+3 terms.  The correct energy locations are better
approximated with the modified Hamiltonian. Fig.
\ref{fig:Single-particle-energy} also shows that neutron subshell
closure only appears with the modified Hamiltonian.  Here, the
ordering is correct but the states are considerably more spread
out compared with experiment.

Let us consider some of the details underlying the
single-particle-like states. The situation for the $1p_{3/2}$ or
``$1p3$'' state in the left panel of Fig.
\ref{fig:Single-particle-energy}, the proton single-particle state
in $^{49}Sc$ with the modified Hamiltonian, is quite interesting.
It appears that this state is mixed over several excited states in
the spectrum.  We can take the strength spread over several states
and construct a centroid for this $1p3$ state by a weighted
average over the states carrying that strength.  Here are the
relevant input ingredients.

The first excited state of $^{49}Sc$ is a ${3/2}^{-}$, as seen in
the second column of Fig. \ref{fig:Sc49}, with about 51\% of the
occupancy of the $1p3$ state.
%To see the 51\%, we add the occupancy of the four single particle states labelled "1   1   3".
Its eigenvalue is $-425.151$ MeV compared to a ground state of
$-428.365$ MeV. The $18th$ state in the $^{49}Sc$ spectrum is also
a ${3/2}^{-}$ with 28\% of the occupancy of the $1p3$ state.  Its
eigenvalue is $-422.803$ MeV. The $24th$ state is also a
${3/2}^{-}$ with 21\% of the occupancy of the $1p3$ state.  Its
eigenvalue is $-422.440$ MeV.

Thus, to a good approximation, the $1p3$ strength is spread over these
three states.  We will identify the weighted average $[0.51 \times (-425.151) +
0.28 \times (-422.803) + 0.21 \times (-422.440)] = -423.79$ as the
centroid of the single particle $1p3$ state which we
then include accordingly in the second column of the figure.

For the proton hole states with the modified Hamiltonian, we
perform a detailed search up to excitation energies of about $14$
MeV in the $^{47}$K spectra. It appears that the $0d_{5/2}$
single-hole state is spread among many states with the largest
observed concentration on the ${5/2}^{+}$ state at $-386.17$ MeV
(13.36 MeV of excitation energy). Here, we find a single $J^{\pi}
= {5/2}^{+}$ state in $^{47}$K with 30 \% $0d_{5/2}$ vacancy and
we assign this state to our $0d_{5/2}$ single-hole state.  Most of
the $0d_{5/2}$ strength, however, was not observed among the
limited number of converged eigenstates.

Let us consider the $^{49}$Ca results with the modified
Hamiltonian in the upper right panel of Fig.
\ref{fig:Single-particle-energy}. The ground state is
approximately a pure $[(1p_{3/2})^1(0f_{7/2})^8]$ configuration.
We note that the spacing for the subshell closure is in good
agreement with experiment while there is a shift of a couple MeV
towards more binding in the model as previously indicated in Fig.
\ref{fig:Ground-state-energy}. A nearly pure $1p_{1/2}$
single-particle state is obtained at 5.235 MeV excitation energy
and an extra low-lying ${7/2}^{-}$ appears with $2p-1h$ character
(see Fig. \ref{fig:Ca49}). Our lowest-lying ${5/2}^{-}$ consists
of $2p-1h$ character relative to subshell closure.

We contrast the modified Hamiltonian's results for the $^{49}$Ca
ground state with those obtained using the {\it ab initio} CD-Bonn where
$[(1p_{3/2})^{4}(1p_{1/2})^{2}(0f_{7/2})^{3}]_{1/2^{-}}$ is the
dominant configuration reflecting again the inadequacies
of the neutron single-particle properties with CD-Bonn.

\subsection{Monopole matrix elements V(ab;T)}

\noindent
The monopole matrix element  is defined by an angular momentum
average of coupled doubly-reduced two-body matrix elements:
\begin{equation}\label{monopole}
V(ab;T) = \frac{\sum_{J} (2J+1) V(abab; JT)} {\sum_{J} (2J+1)} \;
.
\end{equation}
For our NCSM Hamiltonians the ``$V$'' appearing in Eqn.

\ref{monopole} signifies the full 2-body intrinsic-coordinate
Hamiltonian, $T_{rel} + V_{eff}$, except that  we omit the Coulomb
interaction from this analysis.

We examined the monopole character of our initial CD-Bonn
Hamiltonian and we compared it with the monopole features of the
GXPF1 interaction\cite{[HOB04]}. Although the monopole
characteristics were similar, we do not present the detailed
comparisons here due to the ambiguity of the role of the SPE's.
That is, one may shift some Hamiltonian components between SPE's and
two-body matrix elements (TBME's) and this obscures direct
comparisons of a subset of our TBME's with the corresponding subset of GXPF1.
In order to summarize a comparison of the underlying theoretical interactions,
we list in Table 1~a simplified overview of their differences and
similarities.

For a sample comparison of the interactions, we present a small set of two-body
$fp$-shell matrix elements applicable to the present investigation in
Table 2.  For convenience we present two columns of key differences in the
matrix elements:
``diff1'' represents the difference between the G-matrix
and the GXPF1 interaction resulting from adjusting
the G-matrix elements to fit spectra; and,
``diff2'' represents the difference between our
{\it ab initio} $H_{eff}$ and our modified $H_{eff}$.
The scale of the changes from the respective starting
solutions appears comparable, though one of the "diff2"
values reaches -1.2123  MeV.

\subsection{Matrix element correlations}

%Ideally, there is a statement of
% a motivation for a figure, what is displayed and the conclusion(s)
% drawn from that figure.
% In general, there should be at least one critical observation
% emerging from each figure.
% Below, I give some suggestions in order for you to get an idea
% for how to expand the discussions of the figures.
%
% Figure captions should explain the dashed line representing
% the reference correlation line of 45-degrees
%
We present in Figs. \ref{CDB3CDBallME} -
\ref{CDB3GXPF1TrueOff-Diag} the correlations between pairs of
$fp$-shell interaction matrix element sets.  With Fig.
\ref{CDB3CDBallME}, we observe the high degree of correlation
between the 195 matrix elements of our starting Hamiltonian,
CD-Bonn, and our modified Hamiltonian, CD-Bonn + 3 terms.  This
indicates that, for the most part, our Hamiltonian is minimally
modified by the addition of the phenomenological terms. Such a
high correlation is reminiscent of the high correlations seen
between GXPF1 and its starting interaction, the G-matrix
\cite{[HOB04]}. It is interesting to see if certain groups of
matrix elements appear to be more correlated than the others. We
distinguish the diagonal $V(abab; JT)$ matrix elements that
contribute to the monopole by different symbols. Filled circles
stand for $V(abab; JT=0)$ and filled squares stand for $V(abab;
JT=1)$ matrix elements. All the remaining matrix elements $V(abcd;
JT)$ where at least one single-particle-state (sps) of the bra is
different from a sps of the ket are plotted as open circles for
$T=0$ and open squares for $T=1$. We see the filled square
points, that correspond to $V(abab; JT=1)$ matrix elements, are
farther from the linear fit, ranging between 1 MeV and 2 MeV away
from the linear fit line. Therefore, these monopole matrix
elements have received larger corrections than others in the
process of fitting the $A=48$ isotopes.

In the forth paragraph of the same subsection we changed the
sentence "The colors and the symbols used are the same as those in
figures 7 and 8." with "The symbols used are the same as those in
figures 7 and 8."

To see the stronger correlations more clearly, we next choose in
Fig. \ref{CDB3vsCDBafterelimababME} to eliminate all matrix
elements contributing to the monopole.
There are 135 remaining matrix elements out of the 195 total. The
degree of correlation between the 135 matrix elements
significantly improves with much less deviation from the linear
fit. We can see another feature of the correlation by comparing
the linear fit with the $45$-degree line in Fig.
\ref{CDB3vsCDBafterelimababME} (similar pattern seen in Fig.
\ref{CDB3CDBallME}). On the one hand, we see that the CD-Bonn + 3
terms matrix elements are shifted towards greater attraction where
CD-Bonn is already attractive. On the other hand, the CD-Bonn + 3
terms matrix elements are shifted towards greater repulsion where
CD-Bonn is already repulsive. Overall, we observe that the larger
differences between the CD-Bonn and CD-Bonn + 3 terms are coming
from the monopole terms. This seems natural in light of the fact
that the phenomenological terms have the effect of adjusting the
single particle features of the theory towards agreement with
experiment. That is, the monopole terms receive the largest
adjustments as required to achieve the needed single particle
features.

It is then very interesting to observe in Fig. \ref{CDBvsGallME}
the lack of correlation between our starting Hamiltonian, CD-Bonn,
and the G-matrix underlying the GXPF1 interaction. Points are
generally farther away from the fit line than in the correlation
of CD-Bonn + 3 terms with CD-Bonn case. Note that the G-matrix is
a renormalization procedure and the specific results for GXPF1 are
developed from the bare CD-Bonn interaction. This lack of
correlation in Fig. \ref{CDBvsGallME} reflects the major
differences in the underlying effective interaction theories that
are summarized in Table 1.

We now make the same set of comparisons between CD-Bonn and
G-matrix in Figs. \ref{CDBvsGallME} and
\ref{CDBvsGafterelimababME} as we performed in Figs.
\ref{CDB3CDBallME} and \ref{CDB3vsCDBafterelimababME}. The symbols
used are the same as those in Figs. \ref{CDB3CDBallME} and
\ref{CDB3vsCDBafterelimababME}. Comparing Fig. \ref{CDBvsGallME}
and Fig. \ref{CDBvsGafterelimababME} we can see how, after
eliminating the matrix elements that contribute to the monopole,
the correlation is significantly improved with the linear fit in
Fig. \ref{CDBvsGafterelimababME} and it now overlaps well with the
45-degree line. Again, this shows similarities between these two
interactions with the difference arising primarily from the
monopole part.

We can comment further about the comparison presented in Fig.
\ref{CDBvsGallME} by observing that the full Hamiltonian developed
from the G-matrix includes single-particle energy (SPE)
contributions. On the other hand CD-Bonn does not have additional
SPE contributions since those contributions are already included
in the 2-body matrix elements. In fact, those SPE contributions to
our interactions are embedded in the monopole terms. This is one
reason why the correlations improve when we proceed from Fig.
\ref{CDBvsGallME} to Fig. \ref{CDBvsGafterelimababME}, removing
the monopole terms.

Finally, in order to focus as clearly as  possible on the  2-body
interaction effects , we present in Fig. \ref{CDB3GXPF1TrueOff-Diag}
the correlation of matrix elements $V(abcd;
JT)$$(A=48)$ between CD-Bonn+3terms and GXPF1, where we retain
only those that cannot contribute to a single-particle
Hamiltonian.  That is, we eliminate all two-body matrix elements
where at least one single-particle-state (sps) of the bra equals a
sps of the ket. There are 56 remaining two-body matrix elements.  Differences
ranging up to about 3 MeV are observed which should lead to differences
in experimental observables.
Comparisons of spectra and other properties with these
Hamiltonians, as one proceeds further from $A=48$, could shed more
light on their differences.

\section{Conclusions and outlook}

We have presented an initial NCSM investigation of the spectral
properties of the $A=47$ and $A=49$ nuclei that are one nucleon
away from doubly-magic $^{48}$Ca.  We have shown that the NCSM
with a previously introduced modified Hamiltonian produces
spectral properties in reasonable accord with experiment. Shell
closure and single-particle spectral properties are obtained
indicating a path has been opened for multi-shell investigations
of these nuclei within the NCSM.  We are undertaking such
additional investigations. Also, for a better understanding of
various $fp$-shell interactions we made a comparison between our
initial and modified $fp$-shell matrix elements in the harmonic
oscillator basis with the GXPF1 interaction \cite{[HOB04]}. Our
initial and modified NCSM $H_{eff}$ matrix elements in the
$fp$-shell are strongly correlated. We found some evidence
suggesting that significant differences in single-particle
properties may underly some of the distinctions between our
$H_{eff}$ and the GXPF1 interaction. The differences were reduced
when we compared the purely off-diagonal matrix elements in Fig.
\ref{CDB3GXPF1TrueOff-Diag}. Additional applications could reveal
the importance of these distinctions in greater detail.

\section{Acknowledgements}

This work was partly performed under the auspices of
the U. S. Department of Energy by the University of California,
Lawrence Livermore National Laboratory under contract
No. W-7405-Eng-48.
This work was also supported in part by USDOE grants DE-FG02-87ER40371
and DE-FC02-09ER41582, Division of Nuclear Physics, and, in part,
by NSF grant INT0070789.

\newpage
\begin{table}
\begin{center}
\begin{tabular}{|c|c|c|}
\hline
\hline
Hamiltonian Property & G-matrix & NCSM cluster $H_{eff}$ \\
\hline
Oscillator parameter dependence & Yes & Yes\\
\hline
Depends on the choice of P-space & Yes & Yes\\
\hline
Reguires effective multi-nucleon  & & \\ interactions as corrections
 & Yes & Yes \\
\hline
Translationally invariant & No & Yes\\
\hline
Starting energy dependence & Yes & No\\
\hline
Single-particle spectra dependence & Yes & No\\
\hline
$A$-dependence & No & Yes\\
\hline
\hline
\end{tabular}
\end{center}
\caption{Overview of the differences and similarities of the two theoretical
approaches that underlie the Hamiltonians whose matrix elements
are compared in this work.}
\label{t1}
\end{table}

\begin{table}
\begin{center}
\begin{tabular}{|cccccccccccc|}
\hline
\hline
$2j_{a}$ & $2j_{b}$ & $2j_{c}$ & $2j_{d}$ & J & T & G & GXPF1 & diff1
& CD-Bonn & CD-Bonn & diff2\\
  & & & & & & & & & & +3 terms & \\
\hline 7 & 3 & 7 & 3 & 5 & 0 & -2.1167 & -2.8504 & -0.7337 &
-1.0390 &-1.3413 & -0.3023\\
3 & 3 & 5 & 5 & 0 & 1 & -0.5243 & -1.1968 & -0.6725 & -0.6019 &
-1.1129 & -0.5109 \\
7 & 7 & 7 & 7 & 3 & 0 & -0.2309 & -0.8087 & -0.5778 & 0.5597 &
0.5555 & -0.0042 \\
7 & 5 & 7 & 5 & 6 & 0 & -2.3465 & -2.9159 & -0.5693 & -1.3743 &
-1.8599 & -0.4856\\
7 & 5 & 7 & 5 & 5 & 0 & -0.0203 & -0.5845 & -0.5642 & 0.5813 &
0.4117 & -0.1693\\
3 & 1 & 3 & 1 & 2 & 1 & -0.7965 & -0.2822 & 0.5143 & -0.0068 &
-0.4932 & -0.4864\\
7 & 7 & 5 & 5 & 0 & 1 & -1.9095 & -1.3288 & 0.5806 & -2.2586 &
-3.4709 & -1.2123\\
\hline
\hline
\end{tabular}
\end{center}
\caption{Comparison of selected two-body matrix elements $V(abcd;
JT)$ (Mev) $(A=48)$ for which the difference between our
interaction is large. ``diff1'' represents the difference between
GXPF1 and G while ``diff2'' is the difference between
CD-Bonn+3terms and CD-Bonn.} \label{t2}
\end{table}

\end{document}